\documentclass[a4paper,12pt]{article}
\usepackage{amsfonts}
\usepackage{amsmath}
\usepackage{amssymb}
\usepackage{graphicx}
\usepackage{psfrag}
\usepackage{epsfig}

\newcommand{\be}{\begin{equation}}
\newcommand{\ee}{\end{equation}}
\newcommand{\bea}{\begin{eqnarray}}
\newcommand{\eea}{\end{eqnarray}}

\newcommand{\bc}{{\bf c}}
\newcommand{\bh}{{\bf h}}
\newcommand{\bt}{{\bf t}}
\newcommand{\bE}{{\bf E}}
\newcommand{\bJ}{{\bf J}}
\newcommand{\bEt}{{\bf {\tilde E}}}
\newcommand{\bA}{{\bf {\tilde A}}}
\newcommand{\bQ}{{\bf {\hat Q}}}
\def\bQr{{\bf Q}}
\newcommand{\bq}{{\bf q}}

\newcommand{\bS}{{\bf S}}
\newcommand{\bJt}{{\bf {\tilde J}}}
\newcommand{\bIh}{{\bf {\hat I}}}
\newcommand{\bD}{{\bf {\tilde D}}}
\newcommand{\bB}{{\bf {\tilde B}}}
\newcommand{\ba}{{\bf {\tilde a}}}
\newcommand{\bK}{{\bf {\tilde K}}}
\newcommand{\bMt}{{\bf {\tilde M}}}
\newcommand{\bMh}{{\bf {\hat M}}}
\newcommand{\bPh}{{\bf {\hat P}}}
\newcommand{\bXt}{{\bf {\tilde X}}}

\newcommand{\bC}{{\bf C}}
\newcommand{\bCt}{{\bf {\tilde C}}}

\newcommand{\balpha}{\mbox {\boldmath ${\alpha}$}}
\newcommand{\bbeta}{\mbox {\boldmath ${\tilde \beta}$}}
\newcommand{\btau}{\mbox {\boldmath ${\tilde \tau}$}}

\newcommand{\bmu}{\mbox {\boldmath ${\tilde \mu}$}}
\newcommand{\bgamma}{\mbox {\boldmath ${\gamma}$}}
\newcommand{\chat}{{{\bf {\hat c}}}}

\newcommand{\hhat}{{{\bf {\hat h}}}}

\newcommand{\bcar}{{{{\bf \tilde{c}}}}}
\newcommand{\bdar}{{{{\bf \tilde{d}}}}}
\newcommand{\bear}{{{{\bf \tilde{e}}}}}
\newcommand{\bfar}{{{{\bf \tilde{f}}}}}

\newcommand{\sbspc}{\!\!\!\!\!}

\title{Thermodynamics of the multi-component dimerizing hard-sphere Yukawa
mixture in the associative mean spherical approximation.}

\author{S. P. Hlushak, Yu. V. Kalyuzhnyi\\
{\it Institute for Condensed Matter Physics,}\\
{\it Svientsitskoho 1, 79011, Lviv, Ukraine}
}

\date{\today}

\begin{document}

\maketitle

\begin{abstract}

Explicit analytical expressions for Helmholtz free energy, chemical potential, entropy
and pressure of the multi-component dimerizing Yukawa hard-sphere fluid are presented.
These expressions are written in terms of the Blum's scaling parameter $\Gamma$, which
follows from the solution of the associative mean spherical approximation (AMSA) for the
model with factorized Yukawa coefficients. In this case solution of the AMSA reduces to
the solution of only one nonlinear algebraic equation for $\Gamma$. This feature enables
the theory to be used in the description of the thermodynamical properties of associating
fluids with arbitrary number of components, including the limiting case of polydisperse
fluids.

\end{abstract}

\newpage

\section{Introduction}

Much of the progress achieved by the liquid-state integral-equation theories
is due to the availability of the integral-equation approximations (IEA), which are
amenable to the analytical solution for a number of the models of dense fluids
and liquids. Since 1963, when Percus-Yevick approximation for the hard-sphere fluid
was solved analytically \cite{w_prl,thiele}, the analytical solutions were derived
for a large variety of non-trivial Hamiltonian models
(see \cite{mcdonald,kal_rev} and references therein). During the last two
decades substantial efforts have been focused on the development of the analytically
solvable IEA for the models of associating fluids \cite{kal_rev}. Most of these studies
were carried out in the frames of the product-reactant Ornstein-Zernike
approach (PROZA) \cite{w12,w34,wdim,mydim,stelldim,pr1,pr2,pr3,pr4,pr5,pr6,pr7,duda1,duda2}
for the models, which combine hard-sphere
interaction and sticky interaction due to a certain number of the sticky points located
on the surface of each hard sphere. Different versions of the models,
which describe dimerizing \cite{wdim,mydim,stelldim}, polymerizing
\cite{pr1,pr2,pr3,pr4,pr5,pr6} and network forming \cite{pr7,duda1,duda2}
fluids, where investigated . More recently these studies were extended
by adding a van der Waals attraction, modeled by a sum of the Yukawa terms
\cite{Yd1,Yd2,Yp1,Yn1}.
The properties of the corresponding models were studied using the analytical solution
of the associating mean spherical approximation (AMSA) \cite {amsa}.
In the particular case of the multi-component Yukawa dimerizing hard-sphere
fluid with factorizable Yukawa coefficients the Yukawa part of
the solution was reduced to the solution of only one nonlinear algebraic equation
for the Blum's \cite{blum1,blum2} scaling parameter $\Gamma$ \cite{Ginoza,Yd2}.
In the limiting case of complete association, when the system is represented by
the multi-component mixture of Yukawa heteronuclear hard-sphere diatomics,
solution of this equation represents full solution of the AMSA. In this
limit PROZA reduces to the 'proper' site-site theory \cite{KC} due to
Chandler et al. \cite{CSL}.

In this article we extend solution of the AMSA obtained earlier \cite{Yd1,Yd2}
and derive explicit expressions for the thermodynamical properties of the
multi-component Yukawa dimerizing hard-sphere fluid in terms of the $\Gamma$-parameter.

\section{The model}

We consider $M$-component mixture of dimerizing Yukawa hard spheres of
species $i=1,2,\ldots,M$ with diameters $\sigma_i$ and densities $\rho_i$.
Each of the hard spheres has one sticky site placed on a surface. The pair potential
of the model consists of the hard-sphere term, sticky site-site term and Yukawa
term $\Phi_{ij}^{(Y)}(r)$, which was chosen to be of the following form:

\begin{equation}
    \beta\Phi^{(Y)}_{ij}\left( r \right) = -\frac{K_{ij}}{r}e^{-zr}   \label{yukawapotenc}
\end{equation}
where $\beta=1/kT$, $k$ is the Boltzmann constant, $T$ is the absolute temperature
and $r$ is the distance between the centers of the spheres.

\section{Solution of the AMSA}

Solution of the AMSA for the model at hand was obtained earlier
\cite{Yd1,Yd2}
and we shall therefore omit the details here and present only the final
expressions, which are needed in our derivation of the thermodynamics.

AMSA consists of the two-density Ornstein-Zernike equation
\be
\hhat_{ij}(k)=\chat_{ij}(k)+\sum_l\rho_l\chat_{il}(k)\balpha_l\hhat_{lj}(k),
\label{oz}
\ee
supplemented by the MSA-like closure conditions
\be
\bc_{ij}(r)=\bE{K_{ij}\over r} e^{-zr}, \label{amsac}
\hspace{23mm}
r>\sigma_{ij}=(\sigma_i+\sigma_j)/2
\ee
\be
\bh_{ij}(r)=-\bE+{\bt_{ij}\over 2\pi\sigma_{ij}}\delta (r-\sigma_{ij}), \label{amsah}
\hspace{20mm}
r<\sigma_{ij}.
\ee
Here
$\hhat_{ij}(k)$, $\chat_{ij}(k)$, $\bt_{ij}$, $\balpha_i$ and $\bE$ are
the following matrices:
$$
\hhat_{ij}(k)=\begin{pmatrix}
   {\hat c}_{i_0j_0}(k) &{\hat c}_{i_0j_1}(k) \cr
   {\hat c}_{i_1j_0}(k) &{\hat c}_{i_1j_1}(k) \cr
\end{pmatrix},\qquad
\chat_{ij}(k)=
\begin{pmatrix}
   {\hat c}_{i_0j_0}(k) &{\hat c}_{i_0j_1}(k) \cr
   {\hat c}_{i_1j_0}(k) &{\hat c}_{i_1j_1}(k) \cr
\end{pmatrix},
$$
$$
\bt_{ij}=
\begin{pmatrix}
  0 & 0 \cr
          0 & t_{i_1j_1} \cr
\end{pmatrix},
\hspace{10mm}
\balpha_i=
\begin{pmatrix}
    1    & \alpha_i \cr
          \alpha_i &   0      \cr
\end{pmatrix},
\hspace{10mm}
\bE=
\begin{pmatrix}  1 & 0 \cr
          0 & 0 \cr
\end{pmatrix},
$$
where
${\hat h}_{i_\alpha j_\beta}(k)$ and ${\hat c}_{i_\alpha j_\beta}(k)$
are Fourier transforms of the total
$h_{i_\alpha j_\beta}(r)$ and direct $c_{i_\alpha j_\beta}(r)$ correlation 
functions, respectively,
$t_{i_1 j_1}=T_{ij}g_{i_0 j_0}(\sigma_{ij}^+)$, $g_{i_0 j_0}(\sigma_{ij}^+)$ is
the contact value of the radial distribution function
$g_{i_\alpha j_\beta}\left( r \right)
 = h_{i_\alpha j_\beta}\left( r \right) + \delta_{\alpha 0} \delta_{\beta 0}$,
$T_{ij}$ is the parameter which defines the strength of the sticky interaction
and $\alpha_i$ is the fraction of non bonded particles.
Here the lower indices $\alpha$ and $\beta$ denote
the bonding state of the corresponding particle and take the values $0$ (non bonded)
and $1$ (bonded).
Fraction of non bonded particles $\alpha_i$ together
with association strength parameter $T_{ij}$ obey the mass action law (MAL) relation
\be
1=\alpha_i\left( 1+2\sum_j\rho_j\sigma_{ij}\alpha_jt_{i_1 j_1}\right).
\label{density}
\ee

Solution of the AMSA was obtained using Baxter factorization method
\cite{baxter} with the general scheme of the solution based upon the version of
factorization technique developed by H\o{}ye and Blum \cite{hoyeblum1,hoyeblum2}.
According to Baxter \cite{baxter} the OZ equation (\ref{oz}) can be factorized as
\be
\bS_{ij}(|r|)=\bQr_{ij}(r)-\sum_l\rho_l\int dr' \bQr_{il}(r')\balpha_l\bQr_{jl}^T(r'-r),
\label{ozf3}
\ee
\be
\bJ_{ij}(|r|)=\bQr_{ij}(r)+\sum_l\rho_l\int dr'\bJ_{il}(|r'-r|)\balpha_l\bQr_{lj}(r'),
\label{ozf4}
\ee
where $T$ denotes the transpose matrix and the integrals
$
\bS_{ij}(r)=2\pi\int_r^\infty dr'r'{\bf c}_{ij}(r')
$
and
$
\bJ_{ij}(r)=2\pi\int_r^\infty dr'r'{\bf h}_{ij}(r')
$
satisfy the following boundary conditions:
\be
\left\{
\begin{array}{lllll}
\bJ_{ij}(r)&=&\pi r^2\bE+\bJ_{ij},\hspace{10mm}
&r&\le\sigma_{ij}\\
\bS_{ij}(r)&=&\bE{K_{ij}\over z}e^{-zr},
&r&>\sigma_{ij}
\end{array}
\right..
\ee
Here $\bJ_{ij}=\bJ_{ij}(0)$.
From the analysis of the equations (\ref{ozf3}) and (\ref{ozf4}) we get
\cite{Yd2}
\be
\bQr_{ij}(r)=\left[\bq_{ij}(r)+\bt_{ij}\right]\theta(\sigma_{ij}-r)
+\bEt^T\bD_{ij}e^{-zr}, \label{qcap}
\hspace{10mm}
r>\lambda_{ji},
\ee
where $\lambda_{ij}={1\over 2}\left(\sigma_i-\sigma_j\right)$,
$\bD_{ij}$ and $\bEt$ are the row vectors, i.e.
$
\bD_{ij}=\left(D_{i_0 j_0},D_{i_0j_1}\right),
$
$
\bEt=\left(1,0\right).
$
Vector $\bD_{ij}$ satisfies the following relation
\be
{2\pi\over z}\bK_{ij}=\sum_l\rho_l\bD_{il}\balpha_l\bQ_{jl}^T(iz),
\label{kq}
\ee
where
$
\bQ_{ij}(k)=\delta_{ij}\left(\rho_j\balpha_j\right)^{-1}-2\pi\int_{\lambda_{ji}}^\infty
dr\;\bQr_{ij}(r)e^{ikr}
$
and $\bK_{ij}=\left(K_{ij},0\right)$.
Expression for $\bq_{ij}(r)$ in the interval $\lambda_{ji}<r<\sigma_{ij}$ is
\be
\bq_{ij}(r)={1\over 2}{\tilde \bE}^T \bA_j\left(r-\sigma_{ij}\right)\left(r-\lambda_{ji}
\right)+\bEt^T\bbeta_j\left(r-\sigma_{ij}\right)+\bC_{ij}\left(e^{-zr}-e^{-z\sigma_{ij}}
\right).
\label{qr}
\ee
Here
\be
\bbeta_j={\pi \over \Delta}\sigma_j\bEt+{2\pi\over \Delta}\sum_n\bmu_j^{(n)},
\;\;\;
\bA_j={2\pi \over \Delta}\left(\bEt+{1\over 2}\zeta_2\bbeta_j+
\sum_n\bMt_j^{(n)}-\btau_j\right),
\label{ac}
\ee
\be
\bC_{ij}=\sum_l\bgamma_{il}(z)\bEt^T\bD_{lj}-\bEt^T\bD_{ij},
\label{1}
\ee
where
\be
s\bgamma_{ij}(s)=2\pi\rho_j{\bf G}_{ij}(s)\balpha_j,
\hspace{5mm}
\left({\bf G}_{ij}(s)=\int_0^\infty dr'r'{\bf g}_{ij}(r')e^{-sr'}\right),
\label{2}
\ee
\be
\bmu_j^{(n)}=\sum_l\rho_l\bCt_l^\mu(z)\balpha_l\bEt^T\bD_{lj}^{(n)}
e^{-z\sigma_{lj}},
\;\;\;\;
\bMt_j^{(n)}=\sum_l\rho_l\bCt_l^M(z)\balpha_l\bEt^T\bD_{lj}^{(n)}
e^{-z\sigma_{lj}},
\label{mc}
\ee
\be
\bCt_l^\mu(s)=\sum_k\bEt\bgamma^T_{lk}(s)e^{s\sigma_{lk}}s\sigma_k^3
\phi_1(\sigma_ks)+{1\over s^2}\left(1+{1\over 2}s\sigma_l\right)\bEt,
\label{cmu}
\ee
\be
\bCt_l^M(s)=\sum_k\bEt\bgamma^T_{lk}(s)e^{s\lambda_{lk}}\sigma_k^2s
\varphi_1(-s\sigma_k)-{1+s\sigma_l\over s}\bEt.
\label{cm}
\ee
Here $\bgamma_{ij}(z)$ satisfies the following set of the algebraic equations
\be
\sum_lz\bgamma_{il}(z)\bQ_{lj}(iz)=
\bEt\left[\bA_j\left(1+{1\over 2}z\sigma_i\right)+\bbeta_jz\right]
{e^{-z\sigma_{ij}}\over z^2}-\bC_{ij}e^{-2z\sigma_{ij}}+\bt_{ij}e^{-z\sigma_{ij}}
\label{Geq}
\ee
and
\be
\btau_j=\sum_l \rho_l\sigma_l\bEt\balpha_l\bt_{lj},\;\;\;\;
\zeta_m=\sum_l\rho_l\sigma_l^m,\;\;\;\;
\Delta=1-\frac{\pi}{6}\zeta_3,
\label{tau}
\ee
$$
\varphi_1(x)={1-x-e^{-x}\over x^2},\;\;\;
\phi_1(x)={1\over x^3}\left[1-{1\over 2}x-
\left(1+{1\over 2}x\right)e^{-x}\right].
$$
One can see that all coefficients of the factor function $\bQr(r)$ are determined by the set
of unknowns $\bD_{ij}$ and $\bgamma_{ij}(z)$. These unknowns follow from the solution of the
set of equations (\ref{kq}) and (\ref{Geq}).

Substantial simplification of the final algebraic equations representing the solution of the
AMSA occurs in the case of factorizable Yukawa coefficients, i.e. for $K_{ij}=Kd_id_j$.
According to Eq. (\ref{kq}) now $\bD_{ij}$ can be written in the following form
\be
\bD_{ij}=-d_i\ba_je^{{1\over 2}z\sigma_j},
\label{Dd}
\ee
which gives
\be
\bC_{ij}=\left(d_i\bEt^T-{1\over z}\bB_i^T\right)\ba_je^{{1\over 2}z\sigma_j},\label{ccap}
\ee
\be
\bbeta_j={\pi \over \Delta}\sigma_j\bEt+\Delta_1\ba_j,
\label{bbeta}
\ee
\be
\bA_j={2\pi\over \Delta}\left(1+{\pi\over 2\Delta}\zeta_2\sigma_j\right)\bEt+{\pi\over \Delta}P\ba_j-{2\pi\over \Delta}\btau_j.\label{acap}
\ee
Here
\be
\bB_i=z\sum_l\bEt\bgamma^T_{il}(z)d_l,
\label{bcap}
\ee
\be
\Delta_1=-{2\pi\over \Delta}\bEt\sum_l\rho_l\balpha_l\sigma_l^2
\left[\phi_1(z\sigma_l)\sigma_l\bB^T_l e^{{1\over 2}z\sigma_l}+
{1+z\sigma_l/2\over \sigma_l^2z^2}d_l\bEt^Te^{-{1\over 2}z\sigma_l}\right],
\label{bdelta}
\ee
\be
P=\left(\zeta_2-{\Delta\over \pi}z\right)\Delta_1+\bEt\sum_l\rho_l
\balpha_l\sigma_l\left[\varphi_0(z\sigma_l)\sigma_l\bB_l^T
e^{{1\over 2}z\sigma_l}+d_l\bEt^Te^{-{1\over 2}z\sigma_l}\right],
\label{pcap}
\ee
where $\varphi_0(x)=\left(1-e^{-x}\right)/x$. Next, making use of the symmetry property
of the factor function, i.e.
$
\bQr_{ij}(\lambda_{ji})=\bQr^T_{ji}(\lambda_{ij}),
$
we have
\be
\bXt_i^T\ba_j=\ba_i^T\bXt_j,
\label{sym1}
\ee
where
\be
\bXt_i^T=\bEt^T\left(\sigma_i\Delta_1+d_ie^{-{1\over 2}z\sigma_i}\right)
+\sigma_i\bB_i^T\varphi_0(z\sigma_i)e^{{1\over 2}z\sigma_i}.
\label{x1}
\ee
Equation (\ref{sym1}) enables us to introduce scaling parameter $\Gamma$ via the following
relation
\be
\ba_j={2\Gamma\over D}\bXt_j,\label{scaling}
\ee
where $D=\sum_k\rho_k\bXt_k\balpha_k\bXt^T_k$.
Differentiating (\ref{ozf3}) with respect to $r$ and taking the limit
$r\rightarrow 0$ we have
\be
\ba_i={2\over D}\left[-\bEt\Delta_1\left(1+{1\over
2}z\sigma_i\right)-\bB_ie^{{1\over 2}z\sigma_i}
-\sigma_i\bEt\eta^B+\sum_k
\rho_k\bXt_k\balpha_k{\bf t}_{ik}\right],
\label{ba}
\ee
where
\be
\eta^B={\pi\over 2\Delta}\sum_k\rho_k\sigma_k\bXt_k\balpha_k\bEt^T.
\label{etab}
\ee
Now all the unknowns of the problem can be expressed in terms of $\Gamma$, i.e.
\be
X_{i_0}=-\lambda_i-\eta_i\Delta_1-{2\Delta\over \pi}\xi_i\eta^B,
\label{x3}
\ee
\be
X_{i_1}=T_i^{\eta}\Delta_1+
{2\Delta\over \pi}T_i^\xi\eta^B+T_i^\lambda,
\label{x5}
\ee
where
$\;\;\;\pi\sigma_iT_i^y=-2\Delta\xi_i\sum_k\rho_k\alpha_kt_{i_1k_1}y_k,$
($y=\eta,\xi,\lambda$),
$$
\lambda_i=-{d_ie^{-{1\over 2}z\sigma_i}\over 1+\varphi_0(z\sigma_i)\sigma_i\Gamma},
\hspace{7mm}
\eta_i={\sigma_i^3z^2\phi_1(z\sigma_i)\over 1+\varphi_0(z\sigma_i)\sigma_i\Gamma},
\hspace{7mm}
\xi_i={\pi\over 2\Delta}{\sigma_i^2\varphi_0(z\sigma_i)\over 1+\varphi_0(z\sigma_i)\sigma_i\Gamma},
$$
\be
\eta^B={-{\pi\over \Delta}\Theta^\eta\Omega^\lambda+\Theta^\lambda
\left({1\over 2}z^2+{\pi\over \Delta}\Omega^\eta\right)\over
\Theta^\eta\left(2\Gamma+{\pi\over \Delta}\zeta_2+z+2\Omega^\xi\right)+
{\Delta\over \pi}\left(z^2+{2\pi\over \Delta}\Omega^\eta\right)
\left(1-\Theta^\xi\right)},
\label{etaBfin}
\ee
\be
\Delta_1={2\Omega^\lambda\left(\Theta^\xi-1\right)-\left[2\Gamma+
{\pi\over \Delta}\zeta_2+z+2\Omega^\xi\right]\Theta^\lambda\over
\Theta^\eta\left(2\Gamma+{\pi\over \Delta}\zeta_2+z+2\Omega^\xi\right)+
{\Delta\over \pi}\left(z^2+{2\pi\over \Delta}\Omega^\eta\right)
\left(1-\Theta^\xi\right)},
\label{Deltafin}
\ee
\be
\Omega^y=\sum_l\rho_l\left[\alpha_lT_l^y-y_l\left(1-\alpha_l\tau_{l_1}\right)\right],
\hspace{19mm}
\Theta^y=\sum_l\rho_l\sigma_l\left(\alpha_lT_l^y-y_l\right).
\label{omtet}
\ee
Finally, the nonlinear algebraic equation for $\Gamma$, which follows from (\ref{kq}), is
\be
\left(\Gamma\right)^2+z\Gamma+\pi KD=0.
\label{oneeq}
\ee
Full solution of the problem requires solution of the set of equations formed
by equations (\ref{density}) and (\ref{oneeq}). The former equation needs as an input the
contact values of the radial distribution function $g_{i_0j_0}$.
Corresponding expression follows from (\ref{ozf4})
\begin{equation}
g_{i_0j_0}\equiv
g_{i_0 j_0}\left(r\rightarrow \sigma^+_{ij} \right)=\left({1\over\Delta} +
{\xi_{2}\sigma_i\sigma_j\over 4\Delta^{2}\sigma_{ij}}\right)
\exp{\left(\frac{K}{\sigma_{ij}}X^{T}_{i_0}X_{j_0} \right)}. \label{gcontactEXP00}
\end{equation}
Here we have used exponential approximation \cite{blum_bernard}.

\section{Thermodynamics}

Thermodynamic properties of the model at hand will be calculated via the energy route, which
appears to be the most accurate for the MSA-type of the theories. Using standard expression
for the excess internal energy in terms of the radial distribution functions, we have
\begin{equation}
\beta \Delta E^Y = -K\sum_{i}{\rho_{i}d_i\bE \balpha_i \bB^{T}_i}.
\label{deltaE2}
\end{equation}
Before proceeding to Helmholtz free energy calculations we will prove the following two
useful relations
\be
\left[ \frac{\partial \Delta E^Y}{\partial \left(\rho_{i}\alpha_{i}\rho_{j}\alpha_{j}
t_{i_1 j_1}\right) } \right]_{\Gamma=const} = -\frac{K}{\beta}X_{i_0}X_{j_0},
\label{usefulrel2}
\ee
\be
\left[ \frac{\partial \Delta E^Y}{\partial \Gamma} \right]_{\rho_{p}=const} =
- \frac{1}{\pi\beta}\left( {\Gamma}^2 + z\Gamma \right),\label{usefulrel1}
\ee
where $\rho_p$ denotes the set of all products
$\rho_{i}\alpha_{i}\rho_{j}\alpha_{j}t_{i_1 j_1}$.
First of these relations can be derived  by substituting $\bB_i$  from (\ref{x1})
into (\ref{deltaE2}) and differentiating it with respect to
$\rho_{i}\alpha_{i}\rho_{j}\alpha_{j}t_{i_1 j_1}$ with $\Gamma$ held constant.
We will skip these straightforward calculations and proceed to the second of these relations.
To prove relation (\ref{usefulrel1}) we start eliminating $\bXt_{i}$ from (\ref{x1}),
(\ref{ba}) and (\ref{scaling}) to get
$$
    \bB_{i}e^{{1\over 2}z\sigma_i}\left( 1 + \sigma_i\Gamma\varphi_0\left(z\sigma_i\right)
    \right)=
$$
\be
    -\bEt\Delta_{1}\left( 1+{1\over 2}z\sigma_{i}+
    \sigma_i\Gamma \right) -  \bEt\Gamma d_{i}e^{-{1\over 2}z\sigma_{i}} -
    \sigma_{i}\bEt\eta + \sum_{k}\rho_{k}\bXt\balpha_{k}\bt_{ik}.
    \label{useful_auxil1}
\end{equation}
After substituting (\ref{x3}) into the above equation and performing
some lengthy algebra we obtain the following matrix equation for $\bB_{i}$
$$
    \sum_{k}\bB_k \bMh_{ki} =
\left(\left\{ 1+{1\over 2}z\sigma_{i}+\sigma_{i}\Gamma +
    \frac{\pi\sigma_i}{2\Delta}\zeta_2\right\}\bEt -
    \btau_{ik}\right)\frac{2\pi}{\Delta z^2}\sum_{l}\rho_l\left( 1+{1\over 2}z\sigma_l
    \right)d_{l}e^{-{1\over 2}z\sigma_l}
$$
\be
    -\bEt\frac{\pi\sigma_l}{2\Delta}\sum_{k}\rho_{k}\sigma_{k}d_k e^{-{1\over 2}z\sigma_k}
    - \bEt d_i e^{-{1\over 2}z\sigma_i}\Gamma
    +\bJt\left[-\sum_{k}\rho_{k}\alpha_kt_{i_1 j_1}\lambda_k\right.
\label{Bmatrequt}
\ee
$$
    \left.+ \frac{2\pi}{\Delta z^{2}}\sum_{k}\rho_k\left(1+{1\over 2}z\sigma_k\right)d_k
    e^{-{1\over 2}z\sigma_k}
    \sum_{l}\rho_l\alpha_l t_{i_1 j_1}\chi_l -\sum_{k}\rho_{k}\sigma_{k}d_{k}
    e^{-{1\over 2}z\sigma_k}\sum_{l}\rho_l\alpha_{l}t_{i_1 l_1}\xi_l\right]
$$
where $\bJt = \left( 0,\quad 1\right)$,
$\bMh_{ki}  = e^{{1\over 2}z\sigma_k}\left( 1+\varphi_0\left( z\sigma_k \right)\sigma_k
\Gamma \right)\bPh_{ki},
$
\begin{equation}
    \chi_i = \frac{\sigma_i\left\{ 1+{1\over 2}z\sigma_{i}+\sigma_{i}\Gamma + \frac{\pi\sigma_i}{2\Delta}\zeta_2\right\}\varphi_0\left(z\sigma_i\right)}
    {1+\varphi_0\left(z\sigma_i\right)\sigma_i\Gamma},
    \label{chii}
\end{equation}
and $\bPh$ is the Jacobi type of the matrix, i.e.
$\bPh_{ki} =
\delta_{ik}\bIh + \bcar_{k}^T\bdar_{i} + \bear_{k}^T\bfar_{i}.$
Four vectors $\bcar$, $\bdar$, $\bear$ and $\bfar$, that form the
Jacobi matrix, are
$$
\bcar_{i} =  \rho_i \xi_i \left( 1, \quad\alpha_i \right),\;\;\;
\bdar_{i} =  \left(\sigma_i,\quad
-\frac{\sigma_{i}T^{\xi}_i}{\xi_i} \right),\;\;\;
\bear_{i} =  - \frac{2\pi\rho_i}{\Delta z^2} \eta_i \left( 1, \quad\alpha_i \right),
$$
$$
\bfar_{i} =  \left(\left\{ 1+{1\over
2}z\sigma_{i}+\sigma_{i}\Gamma +
\frac{\pi\sigma_i}{2\Delta}\zeta_2\right\}, \quad
-\tau_{i_1}+\sum_{l}\rho_l \alpha_l t_{i_1 l_1}\chi_l \right),
\qquad \bIh=
\begin{pmatrix}
    1    & 0 \cr
    0 &   1      \cr
\end{pmatrix}.
$$
Corresponding equation for $\left[ \frac{\partial\bB_i}{\partial\Gamma} \right]_{\rho_p}$
follows from (\ref{Bmatrequt}) upon its differentiation with respect to $\Gamma$
\begin{equation}
    \sum_{k}\left[ \frac{\partial\bB_k}{\partial\Gamma} \right]_{\rho_p} \bMh_{ki} = -\bXt_i -
    \bJt\left( \frac{2\Delta}{\pi}\sum_{k}^{}\rho_k\alpha_{k}t_{i_1 k_1} \frac{\xi_k}{\sigma_k}X_{k_0} \right).
    \label{partBmatrequat}
\end{equation}
Taking derivative of the both sides of equation (\ref{deltaE2}) with respect to $\Gamma$,
using expression for $\left[ \frac{\partial\bB_i}{\partial\Gamma} \right]_{\rho_p}$, obtained
from the solution of the set of equations (\ref{partBmatrequat}) and taking into account
equation (\ref{oneeq}), we recover relation (\ref{usefulrel1}). Inverse matrix
$\bMh_{ki}^{-1}$  together with $\Gamma$ derivatives, which are used in our
calculations, are given in the Appendix.

We start our derivation of the expression for Helmholtz free energy with the following
standard thermodynamic relation:
\begin{equation}
\frac{\partial}{\partial \beta}\left( \beta \Delta A \right) = \beta \Delta E.
\label{dAisE}
\end{equation}
Integrating this equality by parts and using the fact that all thermodynamic quantities
according AMSA depend only on one parameter $\Gamma$, we have
\begin{equation}
\beta \Delta A^Y = \beta \Delta E^Y - \int_{0}^{\Gamma}d\Gamma' \beta' \frac{d\Delta E^Y}
{d\Gamma'},
\label{bAisbE}
\end{equation}
where $\Delta A^Y$ represent Yukawa contribution to Helmholtz free energy.
We start with the system of dimerizing hard spheres and charge it by the Yukawa
charge $d_i$ up to the current conditions. Full derivative under the integral in (\ref{bAisbE})
can be expressed in terms of the partial derivatives giving
\begin{equation}
    \beta \Delta A^Y = \beta \Delta E^Y - \int_{0}^{\Gamma}d\Gamma' \beta'
    \left[ \frac{\partial\Delta E^Y}{\partial\Gamma'} \right]_{\rho_{p}}
    - \int_{0}^{\Gamma}d\Gamma' \beta' \sum_{ij}
    \left[ \frac{\partial\Delta E^Y}{\partial \rho_{i}\alpha_{i}\rho_{j}\alpha_{j}t_{i_1 j_1}}
    \right]_{\Gamma'} \frac{\partial \rho_{i}\alpha_{i}\rho_{j}\alpha_{j}t_{i_1 j_1}}
    {\partial\Gamma'}.
    \label{bAisbE2}
\end{equation}
Integrating the second integral in (\ref{bAisbE2}) by parts and using
(\ref{usefulrel2}) and (\ref{usefulrel1}), we get
$$
\beta\Delta A^Y = \beta\Delta E^Y + \frac{1}{\pi}\left( \frac{{\Gamma}^3}{3}+
z\frac{{\Gamma}^2}{2} \right)
$$
\be
+ K\sum_{ij}^{}\rho_i\alpha_i\rho_j\alpha_{j}t_{i_1j_1}X_{i_0}X_{j_0}
-\sum_{ij}^{}\int_{0}^{\Gamma}d\Gamma'\rho_i\alpha_i\rho_j\alpha_{j}t_{i_1j_1}
\frac{\partial\left( KX_{i_0}X_{j_0} \right)}{\partial\Gamma'}.
\label{bAisbE3}
\end{equation}
Since
$
\partial t_{j_1 k_1}/\partial\Gamma=t_{j_1 k_1}\partial
    \ln \left(g_{j_0 k_0} \right)/\partial\Gamma
$
and due to the exponential approximation (\ref{gcontactEXP00}) and MAL relation (\ref{density})
it is straightforward to show that
\begin{equation}
\frac{\partial\;\beta A^{MAL}}{\partial\Gamma}=
-\sum_{ij}^{}\rho_i\alpha_i\rho_j\alpha_j t_{i_1 j_1}
\frac{\partial \;KX_{i_0}X_{j_0}}{\partial\Gamma},\label{dAMALis}
\end{equation}
where
\begin{equation}
\beta\Delta A^{MAL}=\sum_{i}^{}\ln\;\alpha_i +\sum_{ij}^{}\rho_i\alpha_i\rho_i\alpha_j
t_{i_1 j_1}.
\label{AMAL}
\end{equation}
The final expression for Helmholtz free energy in excess to Helmholtz free energy of
dimerizing hard-spheres system is obtained combining (\ref{bAisbE3}) and (\ref{dAMALis})
\begin{equation}
\beta\Delta A^Y = \beta\Delta E^Y + \frac{1}{\pi}\left( \frac{{\Gamma}^3}{3}
+z\frac{{\Gamma}^2}{2} \right)
+ K\sum_{ij}^{}\rho_i\alpha_i\rho_j\alpha_{j}t_{i_1j_1}X_{i_0}X_{j_0}
+\beta\Delta A^{MAL} - \beta\Delta A^{MAL}_0,\label{bAisbE4}
\end{equation}
where
\begin{equation}
\beta\Delta A_0^{MAL}=\sum_{i}^{}\ln\;\alpha_i^0
+\sum_{ij}^{}\rho_i\alpha_i^0\rho_j\alpha_j^0\sigma_{ij}t_{i_1 j_1}^0.
\label{MAL0}
\end{equation}
In the case of the reference system represented by the multicomponent hard-sphere mixture we
have
\begin{equation}
\beta\Delta A = \beta\Delta E + \frac{1}{\pi}\left( \frac{{\Gamma}^3}{3}
+z\frac{{\Gamma}^2}{2} \right)
+ K\sum_{ij}^{}\rho_i\alpha_i\rho_j\alpha_{j}t_{i_1j_1}X_{i_0}X_{j_0}
+\beta\Delta A^{MAL},\label{bAamsaExp}
\end{equation}
with $\Delta A$ being Helmholtz free energy in excess to the hard-sphere Helmholtz free energy.
Corresponding expression for the excess entropy $\Delta S$ is found differentiating
(\ref{bAamsaExp}) with respect to the temperature
\begin{equation}
\Delta S=-\frac{k_{B}}{\pi}\left( \frac{{\Gamma}^3}{3}+z\frac{ {\Gamma}^2 }{2} \right) - k_B\beta\Delta A^{MAL}
    - k_B K\sum_{ij}^{}\rho_i\alpha_i\rho_j\alpha_j t_{i_1 j_1}X_{i_0}X_{j_0}.  \label{deltaS}
\end{equation}

Similar as in the earlier studies \cite{blum_bernard,her} the scaling parameter
$\Gamma$ of our theory minimizes the excess Helmholtz free energy
\begin{equation}
    \beta\frac{\partial}{\partial\Gamma}\Delta A^{} = 0.  \label{dAis0}
\end{equation}
Differentiating (\ref{bAamsaExp}) with respect to the density of one of the components $\rho_l$,
we get expression for the chemical potential
\begin{eqnarray}
&&\sbspc\sbspc\sbspc\sbspc\beta\Delta\mu_l =  \beta\left[ \frac{\partial\Delta E}{\partial\rho_l} \right]_{\beta} + \frac{1}{\pi}\left( {\Gamma}^2
    + z\Gamma \right)\left[ \frac{\partial\Gamma}{\partial\rho_l} \right]_{\beta} + \beta\left[ \frac{\partial A^{MAL}}{\partial\rho_l} \right]_{\beta}\nonumber \\
    &&\sbspc\sbspc+ K\sum_{ij}\left( \left[ \frac{\partial\rho_i\alpha_i\rho_j\alpha_j t_{i_1j_1}}{\partial\rho_l} \right]_{\beta}X_{i_0}X_{j_0}
    + \rho_i\alpha_i\rho_j\alpha_j t_{i_1j_1}\left[ \frac{\partial X_{i_0}X_{j_0}}{\partial\rho_l} \right]_{\beta}\right).
    \label{chempotenc}
\end{eqnarray}
Using (\ref{density}) and (\ref{gcontactEXP00}), it can be shown that
$$
    \beta\frac{\partial\Delta A^{MAL}}{\partial\rho_l} = \ln\alpha_l
    - \sum_{ij}\rho_i\alpha_i\rho_j\alpha_j t_{i_1j_1}\left(
    \sigma_{ij}\left[ \frac{\partial \ln{g_{ij}^{HS}\left( \sigma_{ij} \right)}}
    {\partial\rho_l} \right]_{\beta}
+K\left[ \frac{\partial
    \left(X_{i_0}X_{j_0}\right)}{\partial\rho_l} \right]_{\beta}\right)
$$
According to (\ref{usefulrel2}) and (\ref{usefulrel1}) we have
$$
\beta\left[\frac{\partial\Delta E}{\partial\rho_l}\right]_{\beta} = -\frac{\Gamma}{\pi}
    \left(\Gamma+ z \right)
    \left[ \frac{\partial\Gamma}{\partial\rho_l} \right]_{\beta}
    - K\sum_{ij} X_{i_0}X_{j_0}\left[ \frac{\partial\rho_i\alpha_i\rho_j\alpha_jt_{i_1j_1}}
    {\partial\rho_l} \right]_{\beta}
    + \beta\left[ \frac{\partial\Delta E}{\partial\rho_l} \right]_{\rho_p,\Gamma,\beta}.
$$
The latter two expressions, when substituted into (\ref{chempotenc}), yield the following
simple expression for the chemical potential
\begin{equation}
    \beta\Delta\mu_l = \beta\left[ \frac{\partial\Delta E}{\partial\rho_l} \right]_{\rho_p,\Gamma,\beta} + \ln\alpha_l
    -\sum_{ij}\rho_i\alpha_i\rho_j\alpha_jt_{i_1j_1}\left[ \frac{\partial\ln g_{ij}^{HS}\left( \sigma_{ij} \right)}{\partial\rho_l} \right]_{\beta},
    \label{chempotenc2}
\end{equation}
Expression for
$\left[ \frac{\partial\Delta E}{\partial\rho_l} \right]_{\rho_p,\Gamma,\beta}$
was obtained using (\ref{deltaE2}) and (\ref{x1}), it reads
$$
\beta\left[ \frac{\partial\Delta E}{\partial\rho_l} \right]_{\rho_p,\Gamma,\beta} =
-Kd_l\frac{e^{-{z\sigma_l}/{2}}}{\sigma_l\varphi_0\left( z\sigma_l \right)}X_{c0}
$$
$$
- K\sum_i\rho_i d_i\frac{e^{-z\sigma_i/{2}}}{\sigma_l\varphi_0\left( z\sigma_i \right)}
    \left( \left( \alpha_i T^{\eta}_{i}-\eta_i \right)\left[ \frac{\partial\Delta_1}
    {\partial\rho_l} \right]_{\rho_p,\Gamma,\beta}\nonumber \right.
\left. + \frac{2\Delta}{\pi}\left( \alpha_i T^{\xi}_{i}-\xi_i \right)
\left[ \frac{\partial\eta^B}{\partial\rho_l} \right]_{\rho_p,\Gamma,\beta}  \right)
$$
\be
+K\Delta_1 d_l\frac{e^{-{z\sigma_l}/{2}}}{\varphi_0\left( z\sigma_l \right)}+
K d_l ^2\frac{e^{-{z\sigma_l}}}{\sigma_l\varphi_0\left( z\sigma_l \right)}
+\left[ \frac{\partial\Delta_1}{\partial\rho_l} \right]_{\rho_p,\Gamma,\beta}
K\sum_i\rho_i d_i\frac{e^{-{z\sigma_i}/{2}}}{\sigma_i\varphi_0\left( z\sigma_i \right)},
\label{Eexpansion2}
\ee
where derivatives $\left[\frac{\partial\Delta_1}{\partial\rho_l} \right]_{\rho_p,\Gamma,\beta}$ and $\left[\frac{\partial\eta^B}{\partial\rho_l} \right]_{\rho_p,\Gamma,\beta}$ are
given in the Appendix.
Finally for the excess pressure $\Delta P$ one can use the following standard relation:
\be
\beta\Delta P=\beta\sum_i \rho_i \Delta\mu_i - \beta\Delta A,
\ee
Remarkable fact is that chemical potential and pressure are independent of
$\left[ \frac{\partial\Gamma}{\partial\rho_l} \right]_{\beta}$
and thus we don't need to solve any equations to obtain this derivative.

\section{Summary and concluding remarks}

In this paper we consider multi-component dimerizing Yukawa hard-sphere fluid. We present
explicit analytical expressions for Helmholtz free energy,
chemical potential, entropy and pressure of the system
in terms of the Blum's scaling parameter
$\Gamma$, which follows from the solution of the AMSA for the model with factorized Yukawa
coefficients. In the latter case solution of the AMSA reduces to the solution of only one
nonlinear algebraic equation for $\Gamma$. This feature enables the theory to be used
in the description of the structure and thermodynamics of associating fluids with arbitrary
number of components, including the limiting case of polydisperse fluids. We are currently
studying the effects of polydispersity on the phase behavior of the polymer fluid combining
the theory proposed here and dimer thermodynamic perturbation theory for polymers
\cite{chapmanTPTD,KTPTD}.

\section{Appendix}

We present here expressions for the elements of the inverse matrix $\bMh^{-1}$ and
derivatives, which are needed to prove relation (\ref{usefulrel1}) and appear in the
expression for the chemical potential (\ref{chempotenc2})

\begin{eqnarray}
    \left[  \frac{\partial y_i}{\partial\Gamma} \right]_{\rho_p} &=&  - \frac{2\Delta}{\pi\sigma_i}\xi_i y_i,\label{gammaDeriv1}\\
    \left[  \frac{\partial T^{y}_i}{\partial\Gamma} \right]_{\rho_p} &=& -\frac{2\Delta}{\pi\sigma_i}\xi_{i}T^{y}_i
    + \frac{2\Delta}{\pi\sigma_i}\xi_i\sum_{k}\rho_{k}\alpha_{k}t_{i_1 k_1}\frac{2\Delta}{\pi\sigma_k}\xi_k y_k  \label{gammaDeriv2}
\end{eqnarray}
where $y$ takes the values $\xi$,$\eta$ or $\lambda$.
\begin{equation}
    \left[  \frac{\partial \chi_i}{\partial\Gamma} \right]_{\rho_p} = - \frac{2\Delta}{\pi\sigma_i}\xi_i \chi_i + \frac{2\Delta}{\pi}\xi_i.
    \label{partialChi}
\end{equation}

\begin{eqnarray}
    && \sbspc\sbspc\sbspc\sbspc
    \bMh_{ki}^{-1} = \frac{e^{-\frac{z\sigma_i}{2}}}{1+\varphi_0\left( z\sigma_i \right)\sigma_i\Gamma}\left[ \delta_{ik}\bIh
    - \frac{\Delta z^2+2\pi\left( \Omega^{\eta}+\left( \frac{z}{2}+\Gamma+\frac{\pi\zeta_2}{2\Delta} \right)\Theta^{\eta} \right)}{\pi S}\bcar^T_k\bdar_i
    \right. \label{Minv}\\
    &&\sbspc\sbspc\sbspc\sbspc
    \left. + \frac{\Delta z^2\left( 1-\Theta^{\xi} \right)}{{\pi}S}\bear^T_k\bfar_i
    + \frac{2\Theta^{\eta}}{S }\bcar^T_k\bfar_i + \frac{\Delta z^2\left( \Omega^{\xi}+\left( \frac{z}{2}+\Gamma+\frac{\pi\zeta_2}{2\Delta}\right)\Theta^{\xi}\right)}
    {\pi S}\bear^T_k \bdar_i\right]\nonumber,
\end{eqnarray}
where denominator
\begin{equation}
    S = \Theta^\eta\left(2\Gamma+{\pi\over \Delta}\zeta_2+z+2\Omega^\xi\right)+{\Delta\over \pi}\left(z^2+{2\pi\over \Delta}\Omega^\eta\right)
    \left(1-\Theta^\xi\right).  \label{Sdenom}
\end{equation}

\begin{eqnarray}
    \left[\frac{\partial\eta^B}{\partial\rho_l} \right]_{\rho_p,\Gamma,\beta} \sbspc\sbspc &=&
    {\sbspc{\frac{\sigma_l}{2}\left( \frac{\sigma_l^2}{3}\eta^B+X_{c0} \right)
    \left( z^2+\frac{2\pi}{\Delta}\Omega^\eta \right)-\frac{\pi}{2\Delta}\Theta^\eta L_l}\over S},\label{dEtaB_dRho} \\
    \left[\frac{\partial\Delta_1}{\partial\rho_l} \right]_{\rho_p,\Gamma,\beta} \sbspc\sbspc &=&
    {\sbspc{ -\left( 2\Omega^\xi+2\Gamma+z+\frac{\pi\zeta_2}{\Delta} \right)\sigma_l
    \left( \frac{\sigma_l^3}{3}\eta^B + X_{c0} \right) - \left( 1- \Theta^\xi \right)L_l }\over S},
    \label{dDelta1_dRho}
\end{eqnarray}
where
\begin{eqnarray}
    L_l &=& 2\eta^B\sigma_l^2\left( 1+\frac{\pi\zeta_2\sigma_l}{6\Delta} \right)+\frac{\pi\sigma_l^3}{3\Delta}\left( \Omega^\lambda+\Delta_1\Omega^\eta
    +\frac{2\Delta}{\pi}\eta^B\Omega^\xi \right)+2X_{c0}.
    \label{appC}
\end{eqnarray}

\end{document}